\documentclass[preprint,aps,showpacs,nofootinbib,preprintnumbers,amsmath,amssymb]{revtex4-1}
\usepackage{}
\usepackage{epsfig}
\usepackage{subfigure}
\usepackage{dcolumn}
\usepackage{bm}
\usepackage[usenames ,dvipsnames]{xcolor}
\usepackage{slashed}
\usepackage{graphicx,color}

\begin{document}
\title{\Large {\bf Semileptonic $B^-\to f_0(1710\,,1500\,,1370) e^-\bar \nu_e$ decays}}

\author{Y.K. Hsiao$^{1,2}$, C.C. Lih$^{3,1,2}$, and C.Q. Geng$^{2,1}$}
\affiliation{$^{1}$Physics Division, National Center for Theoretical Sciences, Hsinchu, Taiwan 300\\
$^{2}$Department of Physics, National Tsing Hua University, Hsinchu, Taiwan 300\\
$^3$Department of Optometry, Shu-Zen College of Medicine
and Management, Kaohsiung Hsien,Taiwan 452
}
\date{\today}

\begin{abstract}
We study the semileptonic decays of $B^-\to f_0(1710\,,1500\,,1370)  e^-\bar \nu_e$,
in which the three $f_0$ states mix with glueball,  $\bar s s$ and $(\bar u u+\bar d d)/\sqrt 2$ states, respectively.
By averaging the mixings fitted in the literature,
we find that the branching ratios of $B^-\to f_0 e^-\bar \nu_e$
are  $O(10^{-6})$, $O(10^{-6})$ and $O(10^{-5})$, respectively,
which can be simultaneously observed in experiments at $B$ factories.
The large predicted branching rate for $B^-\to f_0(1370) e^-\bar \nu_e$
would provide a clean mode to directly observe the $f_0(1370)$ state.
\end{abstract}

\maketitle
It is believed that some exotic states  with  non-standard internal structures, such as
the four-quark and two-gluon bound states~\cite{Wagner:2012ay}, have been seen already.
For example, the isovector $a_0(980)$ and 
the isodoublet $K^*_0(800)$  can be identified as
$a_0(980) \equiv \bar d u \bar s s$ and $K^*_0(800)\equiv \bar su(\bar u u+\bar d d)$
in the tetraquark (four-quark) picture, instead of
$a_0(980)\equiv \bar d u$ and $K^*_0(800)\equiv \bar s u$ in the standard $\bar q q$ picture.
In addition, since only two of the three isoscalars of
$f_0(1710)$, $f_0(1500)$, and $f_0(1370)$
can be simultaneously fitted into the nonet,
a glueball ($G$) as a multi-gluon bound state can be a solution.
Note that the Lattice QCD (LQCD) calculations predict that
the lightest  glueball of $J^{PC}=0^{++}$ is composed of two gluons
with the mass in the range of 1.5-1.7 GeV  \cite{LQCD-1,LQCD-2}.
These three $f_0$ states clearly mix with the glueball and quark-antiquark states.

Although $f_0(1710)$ or $f_0(1500)$ is taken  
to be mainly a glueball state~\cite{Close1,Close2,XGHe,HY,fitdata1,fitdata2},
the radiative $J/\psi\to f_0(1370)\gamma$ decay via a gluon-rich process
has  not been  observed yet, whereas the other two decays of $J/\psi\to f_0(1710,1500)\gamma$  
are clearly established~\cite{pdg}.
This can be understood from the destructive $G$-$\bar q q$ interference~\cite{Close1,HY}
or simply the weak couplings \cite{BSZou} for the resonant $f_0(1370)\to K\bar K$ ($\pi\pi$)
in  $J/\psi\to K\bar K\gamma$ ($J/\psi\to \pi\pi\gamma$).
Nonetheless, it accords with the doubt of having seen the $f_0(1370)$ state with
direct observations~\cite{Klempt:2007cp,Ochs:2013gi}.
We  note that a resonant scalar state, once identified as $f_0(1370)$~\cite{Li:2011pg,LHCb:2012ae}
in the $\pi\pi$ spectrum of $\bar B^0_s\to J/\psi \pi^+\pi^-$, 
was reexamined to be more like $f_0(1500)$~\cite{Ochs:2013gi}, while
only $f_0(1500)$ is found~\cite{Garmash:2004wa} in the analysis of
 $B^-\to K^+ K^- K^-$.
In addition,  in the $\pi \pi$ spectrum of $D_s^+\to \pi^+\pi^-\pi^+$,
no peak around 1370 MeV is found in the recent investigation \cite{Aubert:2008ao}
and it is not conclusive for  $f_0(1370)$ 
in the $\pi\pi$ spectrum of $J/\psi\to \phi(1020)\pi\pi$~\cite{Ablikim:2004wn} either.
As a result, 
a concrete direct measurement for $f_0(1370)$ is urgently needed.

In this study, we propose to use 
the semileptonic $B^-\to f_0(1370) e^-\bar \nu_e$ decay, arising 
from $b\to u \ell\bar \nu_\ell$ at  quark level, to search for $f_0(1370)$. 
It is interesting to note that,
in contrast with the partly observations 
in the aforementioned weak decays, all three $B^-\to f_0 e^-\bar \nu_e$ decays
can be measured,
providing a new way to simultaneously examine $f_0(1710)$, $f_0(1500)$,
and $f_0(1370)$.
According to the measured branching ratios of $B\to M(\bar n n)e^-\bar \nu_e$~\cite{pdg}
with $M(\bar n n)=\pi^0\,, \eta^{(\prime)}\,, \omega\,,\rho$ and $\bar n n=(\bar u u+\bar d d)/\sqrt 2$,
${\cal B}(B^-\to f_0 e^-\bar \nu_e)$ 
are expected to be of order $10^{-6}-10^{-5}$,
which are accessible to the $B$ factories.
In this report, we  average  the mixings  fitted in the literature~\cite{Close2,HY,fitdata1,fitdata2}
for the three $f_0$ states
to explicitly evaluate  the branching ratios of
 $B^-\to f_0 e^-\bar \nu_e$.


We start with the effective Hamiltonian at quark level, given by
\begin{eqnarray}
{\cal H}(b\to u \ell\bar \nu)&=&\frac{G_F V_{ub}}{\sqrt 2}\;
\bar u\gamma_\mu (1-\gamma_5)b\; \bar \ell\gamma^\mu (1-\gamma_5)\nu\,,
\end{eqnarray}
for the $b\to u$ transition with the recoiled $W$-boson to the lepton pair $\ell \bar \nu$.
The amplitude for $B^-\to f_0^i\,e^-\bar \nu_e$  can be simply factorized as
\begin{eqnarray}\label{amp}
{\cal A}(B^-\to f_0^i\,e^-\bar \nu_e)=\frac{G_F V_{ub}}{\sqrt 2}\alpha_3^i
\langle \bar n n|\bar u\gamma_\mu (1-\gamma_5)b|B^-\rangle \;\bar e\gamma^\mu (1-\gamma_5) \nu_e\;,
\end{eqnarray}
where $\alpha_3^i$ is the coefficient of the mixing state of $\bar n n$  defined in Eq. (\ref{mixingform}).
The matrix element for the $B^-\to \bar n n$ transition is given by
\begin{eqnarray}\label{ff1}
\langle  \bar n n|\bar u\gamma_\mu (1-\gamma_5)b|B^-\rangle
=i\bigg[\bigg(p_\mu-\frac{m_B^2-m_{f(\bar n n)}^2}{q^2}q_\mu\bigg)F_1(q^2)
+\frac{m_B^2-m_{f(\bar n n)}^2}{q^2}q_\mu F_0(q^2)\bigg],
\end{eqnarray}
with $p=p_B-q$ and $q=p_B-p_{\bar n n}=p_e+p_{\bar\nu_e}$,
where the momentum dependences for the form factors $F_{0,1}$ are parameterized in the form of
\begin{eqnarray}\label{ff2}
F(q^2)=\frac{F(0)}{1-a(q^2/m_B^2)+b(q^2/m_B^2)}\,.
\end{eqnarray}
Subsequently, the differential decay width is given by  
\begin{eqnarray}\label{dG}
d\Gamma=\frac{1}{(2\pi)^3}\frac{|\bar {\cal A}|^2}{32M^3_B}dm_{12}^2 dm_{23}^2\,,
\end{eqnarray}
with $m_{12}=p_{f_0}+p_e$, $m_{23}=p_e+p_{\bar \nu_e}$ and
$|\bar {\cal A}|^2$
standing for the amplitude squared derived from Eqs. (\ref{amp}), (\ref{ff1}), and (\ref{ff2}) with
the bar  denoting the summation over lepton spins.

In our numerical analysis, we adopt the PDG~\cite{pdg} to have
$|V_{ub}|=(4.15\pm 0.49)\times 10^{-3}$ and
$(m_{f_0(1710)},\,m_{f_0(1500)},\,m_{f_0(1370)})=(1720,\,1505,\,1350)$ MeV, while
$m_{\bar n n}=1470$ MeV is from Refs.~\cite{HY, Close2}.
The parameters for $F_{0,1}$ shown in Table~\ref{MF}
are calculated in the light-front QCD approach~\cite{LFQM}, 
where we have used the constituent quark masses of $m_{u,d}=0.26\pm0.04$ and $m_b=4.62^{+0.18}_{-0.12}$~GeV and
 the meson decay constants of $f_B$ and $f_\pi$ from the PDG~\cite{pdg}.
We note that our results in Table~\ref{MF} 
are in agreement with those in the perturbative QCD approach \cite{Li:2008tk}.
\begin{table}[t!]
\caption{ The form factors of $B^-\to \bar n n$ at $q^2=0$.}\label{MF}
\begin{tabular}{|c|ccc|}
\hline
$F_{0,1}$&$F(0)$&$a$&$b$\\\hline
$F_0$     &$0.20\pm 0.03$&$0.65^{+0.15}_{-0.05}$&$0.29^{+0.17}_{-0.01}$\\
$F_1$     &$0.20\pm 0.03$&$1.32^{+0.08}_{-0.02}$&$0.64^{+0.11}_{-0.08}$\\\hline
\end{tabular}
\end{table}

Now, we define
\begin{eqnarray}\label{mixingform}
|f_0^i\rangle&=&\alpha_j^i |f_j\rangle\,,
\end{eqnarray}
where 
$f_0^i$ ($i=1,2,3$) stand for $f_0(1710)$, $f_0(1500)$ and $f_0(1370)$, 
$f_j$ ($j=1,2,3$) represent $G$,  $\bar s s$, and $\bar n n=(\bar u u+\bar d d)/\sqrt 2$,
and $\alpha^i_j$ ($i,j=1,2,3$)  are the mixings of a $3\otimes 3$ matrix~\cite{Close1,Close2,XGHe,HY}.

To obtain the mixing matrix $(\alpha_j^i)$, there are two scenarios (I and II) in the literature.
In Scenario I,  $f_0(1500)$ is considered to be the glueball candidate,
such that $f_0(1500)$ with $m_{f_0(1500)}=1505$~MeV has a large mixing to $G$, 
to match with the glueball state with $m_G\simeq 1500$~MeV
in the quenched LQCD calculation \cite{LQCD-1}. 
Here, we take the  mixing matrices  of $(\alpha^i_j)_a$ in Scenario I to be
\begin{eqnarray}\label{s1a}
(\alpha^i_j)_I =
\left (
 \begin{array}{rrr}
 0.36& 0.93& 0.09\\
-0.84& 0.35&-0.41\\
 0.40&-0.07&-0.91\\
 \end{array}
\right )\,,
\left (
 \begin{array}{rrr}
-0.05& 0.95&-0.29\\
 0.80&-0.14&-0.59\\
 0.60& 0.26&  0.75\\
 \end{array}
\right )\,,
\left (
 \begin{array}{rrr}
-0.83&-0.45&-0.33\\
-0.40& 0.89&-0.22\\
-0.39& 0.05& 0.92\\
 \end{array}
\right )\,,\;\
\end{eqnarray}
where $a=1,2,3$ correspond to the three fittings in
Refs.~\cite{Close2,fitdata1,fitdata2}, respectively. 
We remark that
although $|\alpha_1^2|$~\cite{fitdata2} in the third matrix of Eq.~(\ref{s1a})
related to $G$ is  small,
it is still reasonable to have the $a=3$ case in Scenario I 
as $m_G$ is fitted to be 1580 MeV, which is close to the quenched LQCD value.
We note that the  signs of $\alpha_j^i$ vary 
due to the different theoretical inputs.
In this study, we shall take 
 the absolute values,
$|\alpha^i_j|$, to represent the magnitudes of the mixings and
 average them in terms of  
\begin{eqnarray}\label{avg}
\bar{\alpha}^i_j=\frac{\Sigma_{a=1}^3 |\alpha_j^i|_a}{3}\,,\;\;
\Delta \bar{\alpha}^i_j=\sqrt{\frac{\Sigma_{a=1}^3 (\bar{\alpha}^i_j-|{\alpha}^i_j|_a)^2}{3}}\,,
\end{eqnarray}
where $\bar{\alpha}^i_j$ is  the central value of each averaged absolute mixing and
$\Delta \bar{\alpha}^i_j$  reflects the deviation among the fittings. 
As a result, 
from Eq.~(\ref{s1a}) we obtain
\begin{eqnarray}\label{s1}
(\bar{\alpha}^i_j)_I =
\left (
 \begin{array}{ccc}
0.41\pm 0.32&0.78\pm 0.23&0.24\pm 0.10\\
0.68\pm 0.20&0.46\pm 0.32&0.41\pm 0.15\\
0.46\pm 0.10&0.13\pm 0.10&0.86\pm 0.08
 \end{array}
\right )\,.
\end{eqnarray}

Scenario II prefers $f_0(1710)$ instead of $f_0(1500)$ as a glueball state with
$m_G\simeq 1700$~MeV, also predicted by the unquenched LQCD~\cite{LQCD-2}. 
In this scenario, the fitted values for $\alpha^i_j$ in Refs.~\cite{HY,fitdata1,fitdata2} are given by
\begin{eqnarray}\label{s2a}
(\alpha^i_j)_{II} =
\left (
 \begin{array}{rrr}
 0.93& 0.18& 0.32\\
 0.03& 0.84&-0.54\\
-0.36& 0.51& 0.78\\
 \end{array}
\right ),
\left (
 \begin{array}{rrr}
-0.96& 0.17&-0.23\\
      0&-0.82& 0.57\\
 0.29& 0.55&  0.79\\
 \end{array}
\right ),
\left (
 \begin{array}{rrr}
-0.99&-0.05&-0.04\\
-0.03& 0.90&-0.42\\
-0.05& 0.41&  0.90\\
 \end{array}
\right ),\;\
\end{eqnarray}
respectively.
Note that  the three $|\alpha^1_1|$ values in Eq.~({\ref{s2a})
are consistently bigger than 0.9, indicating $f_0(1710)$ to be mainly $G$.
Similarly, from Eq.~(\ref{s2a}) we get
\begin{eqnarray}\label{s2}
(\bar{\alpha}^i_j)_{II} =
\left (
 \begin{array}{ccc}
0.96\pm 0.02&0.13\pm 0.06&0.20\pm 0.12\\
0.02\pm 0.01&0.85\pm 0.03&0.51\pm 0.06\\
0.23\pm 0.13&0.49\pm 0.06&0.82\pm 0.05
 \end{array}
\right )\,.
\end{eqnarray}
Consequently, from  the two scenarios in Eqs.~(\ref{s1}) and (\ref{s2}),
 the branching ratios of $B^-\to f_0(1710,1500,1370) e^-\bar \nu_e$ can be calculated
 based on Eqs.~(\ref{amp})-(\ref{dG}).
Our results are shown in Table~\ref{branchingratios},
where the uncertainties come from $|\alpha_3^i|$, $|V_{ub}|$, and $F_{0,1}$, respectively.
 
With the mixing matrix elements in Eqs.~(\ref{s1}) and  (\ref{s2}),
we are able to specifically study the productions of the  three $f_0$  states before the measurements. 
For example, we  find that  
${\cal B}(B^-\to f_0(1370) e^-\bar \nu_e)$  is about $2.57(2.33)\times 10^{-5}$ 
in Scenario I (II).
Besides, ${\cal B}(B^-\to f_0(1710) e^-\bar \nu_e)$ and ${\cal B}(B^-\to f_0(1500) e^-\bar \nu_e)$
in the two scenarios are predicted to be of order $10^{-6}$.
Since ${\cal B}(B^-\to G e^-\bar \nu_e)$  has been
demonstrated to be as small as $1.1\times 10^{-6}$ \cite{Wang:2009rc}, 
where the magnitude of the uncertainty is as large as the central value,
its contribution to ${\cal B}(B^-\to f_0 e^-\bar \nu_e)$ can be negligible.
The only exception is that, due to the largest $|\alpha^1_1|=0.96$ for Scenario II in Eq. (\ref{s2}),
${\cal B}(B^-\to f_0(1710) e^-\bar \nu_e) \simeq1.0\times 10^{-6}$ 
from the $B\to G$ transition, which is compatible to 
${\cal B}(B^-\to f_0(1710) e^-\bar \nu_e) \simeq1.4\times 10^{-6}$ from the $B\to \bar n n$ transition.
With the branching ratios to be of order $10^{-6}-10^{-5}$,
it is possible to measure the three modes simultaneously.
This will improve the knowledge of the mixing matrix as well as the glueball.
\begin{table}[h!]
\caption{ The branching ratios of $B^-\to f_0(1710,1500,1370) e^-\bar \nu_e$ decays
with the uncertainties corresponding to those in $|\alpha_3^i|$, $|V_{ub}|$, and $F_{0,1}$, respectively.
}\label{branchingratios}
\begin{tabular}{|c|c|}
\hline
mode               &Scenario I\\\hline
$f_0(1710)$     &$(1.96^{+1.97+0.49+0.65}_{-1.29-0.43-0.52})\times 10^{-6}$\\
$f_0(1500)$     &$(5.89^{+5.09+1.47+1.81}_{-3.52-1.31-1.58})\times 10^{-6}$\\
$f_0(1370)$     &$(2.57^{+0.50+0.64+0.83}_{-0.45-0.57-0.67})\times 10^{-5}$\\\hline
\end{tabular}\\
\begin{tabular}{|c|c|}
\hline
mode               &Scenario II\\\hline
$f_0(1710)$     &$(1.36^{+2.12+0.34+0.47}_{-1.14-0.30-0.33})\times 10^{-6}$\\
$f_0(1500)$     &$(9.11^{+2.27+2.28+2.79}_{-2.02-2.02-2.44})\times 10^{-6}$\\
$f_0(1370)$     &$(2.33^{+0.29+0.58+0.75}_{-0.28-0.52-0.60})\times 10^{-5}$\\\hline
\end{tabular}
\end{table}

In sum, 
by averaging the mixings of $|\alpha^i_j|$, fitted 
from the most recent studies in the literature, we have found that
${\cal B}(B^-\to f_0(1370) e^-\bar \nu_e)$
are around $2.6$ and $2.3\times 10^{-5}$ in  Scenarios I and II, respectively.
This decay mode is promising to be measured in the $B$ factories,
which would resolve the doubt  for the existence of $f_0(1370)$.
In addition, we have also shown that
${\cal B}(B^-\to f_0(1710) e^-\bar \nu_e)$ and ${\cal B}(B^-\to f_0(1500) e^-\bar \nu_e)$
are of order $10^{-6}$.
The measurements of these three modes will provide us with some useful information about 
the three $f_0$ states.

\section*{ACKNOWLEDGMENTS}
This work was partially supported by National Center for Theoretical
Sciences, SZL10204006, National Science Council  
 (NSC-102-2112-M-471-001-MY3 and
 NSC-101-2112-M-007-006-MY3) and National Tsing Hua
University (103N2724E1).

\end{document}